\documentclass[pra,reprint,superscriptaddress,floatfix]{revtex4-1}

\usepackage{graphicx}   
\usepackage{subcaption}
\usepackage{caption} 
\usepackage{amsmath}    
\usepackage{amssymb}    
\usepackage{braket}     
\usepackage{bm}         
\usepackage{tikz}       
\usetikzlibrary{decorations.markings} 
\usepackage{comment}
\usepackage{etoolbox}
\usepackage{hyperref}
\hypersetup{
    colorlinks=true,
    linkcolor=blue,
    citecolor=black,
    urlcolor=black
}
\usepackage{cleveref}
\usepackage{url}
\usepackage{xcolor} 

\crefname{section}{Section}{Sections}
\Crefname{section}{Section}{Sections}
\crefname{equation}{Eq.}{Eqs.}
\Crefname{equation}{Equation}{Equations}
\crefname{figure}{Fig.}{Figs.}
\Crefname{figure}{Figure}{Figures.}
\crefname{table}{Table}{Tables}
\Crefname{table}{Table}{Tables}
\crefname{appendix}{Appendix}{Appendices}
\Crefname{appendix}{Appendix}{Appendices}

\newcommand{\Tr}{\operatorname{Tr}}

\begin{document}

\title{Asymptotic TCL4 Generator for the Spin-Boson Model: Analytical Derivation and Benchmarking}

\author{Prem Kumar}
\email{premkr@imsc.res.in}
\affiliation{Optics and Quantum Information Group, The Institute of Mathematical Sciences, C.I.T. Campus, Taramani, Chennai 600113, India.}
\affiliation{Homi Bhabha National Institute, Training School Complex, Anushakti Nagar, Mumbai 400094, India}

\author{K. P. Athulya}
\email{athulyakp@imsc.res.in}
\affiliation{Optics and Quantum Information Group, The Institute of Mathematical Sciences, C.I.T. Campus, Taramani, Chennai 600113, India.}
\affiliation{Homi Bhabha National Institute, Training School Complex, Anushakti Nagar, Mumbai 400094, India}

\author{Sibasish Ghosh}
\email{sibasish@imsc.res.in}
\affiliation{Optics and Quantum Information Group, The Institute of Mathematical Sciences, C.I.T. Campus, Taramani, Chennai 600113, India.}
\affiliation{Homi Bhabha National Institute, Training School Complex, Anushakti Nagar, Mumbai 400094, India}

\begin{abstract}
The spin-boson model is a widely used model for understanding the properties of a two-level open quantum system. Accurately describing its dynamics often requires going beyond the weak system-environment coupling approximation. 
However, calculating the higher-order generators of such a dynamics, with a system-environment coupling that is not too weak, has been known to be challenging, both numerically and analytically.
This work presents the analytical derivation of the complete fourth-order time-convolutionless (TCL) generator for a generic spin-boson model, accurate up to 4th order in the system-environment coupling parameter, under the assumption that the environmental spectral density is an odd function of frequency.
In the case of a semiconductor double-quantum-dot system, our results reveal corrections to the dynamics that may become physically significant in some parameter regimes.
Furthermore, we report that the widely used second-order TCL master equation tends to overestimate the non-Markovianity of a dynamics over a large parameter regime.
Within the regime of its applicability, our results provide a computational advantage over numerically exact techniques.
The accuracy of the fourth-order TCL generator is rigorously benchmarked against specialized analytical calculations done for the Ohmic spectral density with Drude cutoff and against the numerically exact Hierarchical Equations of Motion technique.
\end{abstract}

\maketitle

\section{Introduction}
\label{sec:introduction}
The spin-boson model (SBM) serves as a fundamental paradigm in the study of open quantum systems, finding applications in diverse fields ranging from quantum chemistry and biology to quantum thermodynamics and information processing \cite{gilmore2006criteria, Gilmore_2005, gilmore2008quantum, porras2008mesoscopic, cheche2001dynamics, merkli2013electron, magazzu2018probing}. Understanding the dynamics of the SBM is hence of great importance. Numerically exact techniques to solve for the open system dynamics, like Hierarchical Equations of Motion (HEOM) \cite{tanimura2020numerically} and Tensor Networks \cite{strathearn2018efficient}, have therefore proven to be crucial in the study of such systems. In the weak coupling limit, the Time-Convolutionless (TCL) Master Equation (ME) provides a perturbative approach in the system-environment (SE) coupling strength to solve for the open system dynamics, and can provide great numerical and conceptual advantage over these numerically exact techniques \cite{chen2025benchmarking, breuer2001time, breuer2002theory}.

The TCL2-ME (second-order TCL), equivalent to the Bloch-Redfield ME, has been widely adopted in the literature. However, the TCL2-ME has obvious limitations at higher SE coupling because the corresponding dynamics has $O(\lambda^2)$ errors, where $\lambda$ denotes a dimensionless SE coupling parameter. Consequently, it predicts steady-state (SS) populations correctly only up to zeroth-order ($O(1)$) in $\lambda$, and coherences up to $O(\lambda^2)$ \cite{PhysRevB.71.035318, thingna2012generalized, PhysRevLett.121.070401,thingna2012generalized}. To capture higher-order corrections, that is, the $O(\lambda^2)$ corrections to SS populations and $O(\lambda^4)$ corrections to SS coherences, one must resort to higher-order MEs like the TCL4-ME.

Both numerical and analytical evaluation of the higher order TCL$2n$ generator is considered analytically cumbersome because of the existence of $2n-1$ time-like and $n$ frequency-like integrals, although some progress has been made to calculate these higher-order TCL generators \cite{PhysRevB.71.035318, crowder2024invalidation, PhysRevB.111.115423, PhysRevA.111.042214, chen2025benchmarking}. For example, Crowder et al. \cite{crowder2024invalidation} have greatly reduced the numerical cost of evaluating the TCL4 generator for an arbitrary $N$-dimensional system.

In our previous work \cite{PhysRevB.111.115423}, two specific matrix elements of the asymptotic TCL4 generator were analytically calculated for a generic SBM for all odd spectral densities.
This allowed us to determine the $O(\lambda^2)$ corrections to the SS populations and demonstrate their equivalence with the corresponding terms in the generalized Gibbs state of the system.

The present work significantly extends our earlier results by presenting the derivation of the full asymptotic TCL4 generator for the generic SBM, subject to the same assumption of an odd spectral density. We have been able to reduce the final expression for the coefficients of the TCL4 generator to, at most, a double integral over frequency variables, which, without knowing the specific form of the spectral density, cannot be further simplified. This has greatly facilitated both symbolic analysis and numerical evaluation of the TCL4  dynamics and SS. For example, our results have enabled us to analytically evaluate the 4th-order corrections to the SS coherences of the system.

As an application, we study the corrections that TCL4 provides to the dynamics and SS of a semiconductor double-quantum-dot (DQD) system. We also study the corrections to non-Markovianity induced by TCL4 in the dynamics of TCL2-ME, quantified by the BLP measure. We report that for the majority of the parameter regime considered, TCL2 tends to overestimate the non-Markovianity of a dynamics.
The accuracy of the TCL4 generator is rigorously benchmarked against specialized analytical calculations done for the Ohmic spectral density with Drude cutoff and against the HEOM.
Finally, all the results mentioned in this work, including symbolic derivations, final expressions, numerical evaluation for specific parameters, and plots, etc, in their complete form, have been made publicly available as the accompanying supplementary material \cite{SupplementaryMaterial}.

The remainder of this paper is organised as follows. \Cref{sec:theoretical_framework} introduces the SBM and the TCL ME formalism and notations. \Cref{sec_results} presents our main results. \Cref{sec_numerical_verification} details the numerical checks and benchmarks performed to validate our TCL4 generator. Finally, \cref{sec_conclusion} provides discussion, conclusion, and future directions.

\section{Theoretical Framework}
\label{sec:theoretical_framework}

This section introduces the basic notations, which are reproduced here from \cite{PhysRevB.111.115423} for the convenience of the readers.

\subsection{The Spin-Boson Model}
\label{sec_SBM_model} 

The spin-boson model describes a two-level open quantum system interacting with a large number of uncoupled thermal Bosonic oscillators. The total Hamiltonian, $\hat{H}_{SE}$, 
\begin{align}
    \hat{H}_{SE} &= \hat{H}_S + \hat{H}_E + \hat{H}_I \label{eqn_H_SE},
\end{align} 
is composed of three parts: the system ($\hat{H}_S$), environment ($\hat{H}_E$) and the interaction Hamiltonian ($\hat{H}_I$).
The SE interaction is assumed to have the form:
\begin{align}\label{eqn_H_interaction}
    \hat{H}_I &= \lambda \hat{A} \otimes \hat{B}.
\end{align} 
Here, $\hat{A}$ and $\hat{B}$ are the system and the environment operators, respectively.
For the current work, we assume the following specific forms for all of these operators \cite{breuer2002theory, weiss2012quantum}
\begin{align}\label{eqn_sb_hamiltonian}
    \hat{H}_S &\equiv \frac{\Omega}{2} \hat{\sigma}_3,\\
    \hat{H}_E &= \sum_k\left[\frac{\hat{p}_k^2}{2 m_k}+\frac{1}{2} m_k \omega_k^2 \hat{q}_k^2\right],\\
    \hat{A} &= a_3 \hat{\sigma}_3 - a_1 \hat{\sigma}_1, \label{eqn_spin_boson_A}\\
    \hat{B} &= \sum_k c_k \hat{q}_k. \label{eqn_spin_boson_B}
\end{align}
Here $\hat{\sigma}_i$ represents the Pauli matrices. For the $k$-th environment oscillator, $\hat{p}_k$ and $\hat{q}_k$ are the momentum and position operators, respectively, with $m_k$ being its mass and $\omega_k$ its frequency.
The term $c_k$ represents the strength of the coupling between the system and the $k$-th oscillator in the environment.
Throughout this work, unless otherwise mentioned, we have chosen $a_1 = \sin(\theta)$ and $a_3 = \cos(\theta)$. 

The corresponding spectral density is defined as
\begin{align}
    J(\omega) &= \sum_k \frac{c_k^2}{m_k \omega_k}\delta(\omega - \omega_k).
\end{align} 
The dynamics of the system is determined by the two-point correlation functions of the environment operator $\hat{B}$, defined as
\begin{align}
    \eta(t) &\equiv \text{Im}\{\braket{\hat{B}, \hat{B}(-t)}\},\\
    \nu(t) &\equiv \text{Re}\{\braket{\hat{B}, \hat{B}(-t)}\},
\end{align} 
These correlation functions can be calculated directly from the spectral density as \cite{breuer2002theory}
\begin{align}
    \eta(t) &= -\int_0^\infty d\omega J(\omega ) \sin (t \omega ),\label{eqn_eta_expr}\\
    \nu(t) &= \int_0^\infty d\omega f(\omega ) \cos (t \omega ). \label{eqn_nut_expr}
\end{align} 
Here we have defined
\begin{align}\label{eqn_definition_of_f_omega}
    f(\omega) &\equiv J(\omega) \coth(\beta \omega/2),
\end{align} 
where $\beta$ is the inverse temperature of the environment.

\subsection{TCL Master Equation}
To facilitate the master equation calculations, we adopt a vectorized representation for the system's quantum state. The components of this state vector are defined by the expectation values of the Pauli operators, as follows:
\begin{align}\label{eqn_vectorization_of_qubit_state}
    v_i(t) &= \Tr \{ \hat{\sigma}_i \hat{\rho}_S(t) \},
\end{align}
where $\hat{\sigma}_0$ represents the identity matrix.

The TCL-ME offers a perturbative approach, in the SE coupling strength $\lambda$, to the open quantum system dynamics \cite{breuer2002theory}. In the context of the SBM, TCL-ME can be expressed as:
\begin{align}
    \dot{\vec{v}}(t) = \sum_{n=0}^\infty \lambda^{2n} F^{(2n)}(t) \vec{v}(t),
\end{align}
The term $F^{(2n)}(t)$ is a $4 \times 4$ matrix with elements $F_{ij}^{(2n)}(t)$, which serves as the generator for the system's dynamics at the $2n$-th order. The lowest-order term, $F^{(0)}(t)$, corresponds to the free Hamiltonian evolution. The second-order generator, $F^{(2)}(t)$, yields the well-known Bloch-Redfield ME. Higher-order terms ($n>1$) provide subsequent corrections to the Bloch-Redfield ME. Throughout this work, we adopt natural units where $\hbar = 1$ and $k_B = 1$.

We now define the asymptotic, or long-time, limit of these generators as:
\begin{align}
    F_{ij}^{(2n)} &\equiv \lim_{t \to \infty} F_{ij}^{(2n)}(t).
\end{align}
The coefficients $F_{ij}^{(2n)}$ can in principle be determined systematically order by order using the TCL projection operator method \cite{breuer2001time, breuer2002theory, PhysRevB.111.115423, SupplementaryMaterial}.

\section{Results}
\label{sec_results}
\subsection{The Full Asymptotic TCL4 Generator}

The primary theoretical contribution of this work is the analytical derivation of all the matrix elements of the asymptotic fourth-order generator, $F^{(4)}$, for the generic SBM under the assumption that $J(\omega)$ is an odd function (i.e., $J(\omega) = - J(-\omega)$).
The generic TCL4 matrix elements initially have an integral form which can be illustrated by taking the specific example of $F_{33}^{(4)}$ as follows \cite{breuer2001time, PhysRevB.111.115423}
\begin{widetext}
\begin{align}
    F_{33}&^{(4)} = \lim_{t\to \infty} 16 a_1^2  \int_0^{t} dt_1 \int_0^{t_1} dt_2  \int_0^{t_2} dt_3 
    \bigg\{ a_3^2 \bigg[\eta \left(t_1-t_2\right) \eta \left(t-t_3\right) \left(\cos \left(\left(t-t_2\right) \Omega \right)-\cos \left(\left(t-t_1\right) \Omega \right)\right)\notag\\
    &+ \eta \left(t-t_2\right) \eta \left(t_1-t_3\right) \left(\cos \left(\left(t-t_2\right) \Omega \right)-\cos \left(\left(t-t_1\right) \Omega \right)\right)+ \nu \left(t_1-t_2\right) \nu \left(t-t_3\right) \left(\cos \left(\left(t-t_3\right) \Omega \right)-\cos \left(\left(t-t_2\right) \Omega \right)\right)\bigg]\notag\\
    &+ a_1^2 \bigg[\nu \left(t_1-t_2\right) \nu \left(t-t_3\right) \sin \left(\left(t-t_2\right) \Omega \right) \sin \left(\left(t_1-t_3\right) \Omega \right)+\nu \left(t-t_2\right) \nu \left(t_1-t_3\right) \sin \left(\left(t_1-t_2\right) \Omega \right) \sin \left(\left(t-t_3\right) \Omega \right)\bigg]
    \bigg\}.
    \label{eqn_a334_triple_integral}
\end{align}
\end{widetext}

To calculate the TCL4 coefficients for a general odd spectral density, we must solve a five-dimensional integral. This integral involves three time variables ($t_1$, $t_2$, and $t_3$) and two frequency variables ($\omega_1$ and $\omega_2$, which are part of the functions $\eta(t)$ and $\nu(t)$). We first evaluate the triple integral over the time variables, a task readily handled by a computer algebra system such as \textit{Mathematica}. Following this, we carefully calculate the remaining double integral over the frequency variables in the large $t$ limit.

Note that the interchange of ordering of integrals is not allowed for a general integral, as Fubini's theorem may be violated \cite{thomas1996calculus}. But in the present case, we have numerically verified the validity of the interchange of the frequency and time integrals in \cref{eqn_a334_triple_integral} for the specific choice of Ohmic spectral density with Drude cutoff (see section \cref{sec_drude_verification}). We emphasize this because we did encounter violation of Fubini's theorem while evaluating the remaining frequency integrals, as we have reported previously \cite{PhysRevB.111.115423}.

In our previous work \cite{PhysRevB.111.115423}, the elements $F_{30}^{(4)}$ and $F_{33}^{(4)}$ were reduced to a single integral over a frequency variable $\omega$ (for $\omega$ going from $-\infty$ to $+\infty$), which cannot be further reduced without knowing the form of the spectral density $J(\omega)$. These two matrix elements of $F^{(4)}$ enabled us to analytically calculate the $O(\lambda^2)$ corrections to the SS populations and also acted as a proof of principle for the evaluation of these integrals using \textit{Mathematica}. We now move to the evaluation of the full TCL4 generator for the general SBM.

General properties of the TCL$2n$ generator matrices include $F_{0k}^{(2n)}=0$ for all $k$ due to trace preservation condition ($\dot{v}_0(t)=0$). Furthermore, it can be shown that symmetries related to the structure of $\hat{A}$ leads to constraints $a_3 F_{3k}^{(2n)}(t) = a_1 F_{1k}^{(2n)}(t)$ for all $n\geq1$ \cite{PhysRevB.111.115423}. Overall, the $F^{(4)}$ generator matrix ($F^{(2n)}$ for $n\geq 1$, for that matter) has exactly 8 independent parameters, which we have evaluated in this work.

Given the complex nature of the analytical expressions for the matrix elements of $F^{(4)}$, they are not reproduced in their entirety within this paper. Please see the accompanying supplementary materials for the full derivation and the final results in electronic form \cite{SupplementaryMaterial}.

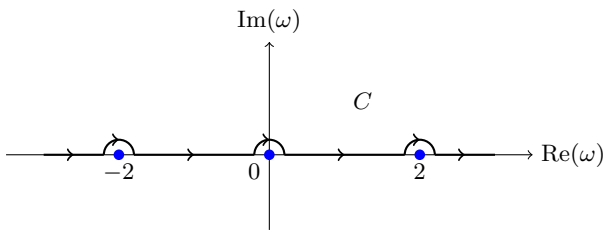
\begin{figure}[htbp]
    \centering
    \begin{tikzpicture}[scale=1, decoration={markings, mark=at position 0.5 with {\arrow{>}}}]
        \draw[->] (-3.5,0) -- (3.5,0) node[right] {\(\text{Re}(\omega)\)};
        \draw[->] (0,-1) -- (0,1.5) node[above] {\(\text{Im}(\omega)\)};

        \def\eps{0.2}
        \def\zeropole{0}

        \draw[thick, postaction={decorate}] (-3,0) -- (-2 - \eps,0);

        \draw[thick, postaction={decorate}] (-2 - \eps,0) arc (180:0:\eps);

        \draw[thick, postaction={decorate}] (-2 + \eps,0) -- (\zeropole - \eps,0);

        \draw[thick, postaction={decorate}] (\zeropole - \eps,0) arc (180:0:\eps);

        \draw[thick, postaction={decorate}] (\zeropole + \eps,0) -- (2 - \eps,0);

        \draw[thick, postaction={decorate}] (2 - \eps,0) arc (180:0:\eps);

        \draw[thick, postaction={decorate}] (2 + \eps,0) -- (3,0);

        \fill[blue] (-2,0) circle (2pt);
        \fill[blue] (\zeropole,0) circle (2pt);
        \fill[blue] (2,0) circle (2pt);

        \node[below] at (-2,0) {$-2$};
        \node[below] at (-0.2,0) {$0$};
        \node[below] at (2,0) {$2$};

        \node[above right] at (1,0.5) {$C$};
    \end{tikzpicture}
    \caption{Pictorial illustration of the contour employed for the $\omega$-integrals in the expression for our TCL4 results (see \cref{eqn_symbolic_tcl4_solution}). The contour is chosen to run above every pole on the real axis, depicted here as blue dots.
}
    \label{fig_contour_avoiding_poles}
\end{figure}

We note that a single universal \textit{Mathematica} function was written to evaluate all the matrix coefficients of $F^{(4)}$. The methodology mirrors that used for $F_{30}^{(4)}$ and $F_{33}^{(4)}$ in our previous work \cite{PhysRevB.111.115423}, albeit streamlined for automated evaluation of the generic matrix element for our convenience. The final result generically looks like
\begin{align}\label{eqn_symbolic_tcl4_solution}
    F_{ij}^{(4)} &= h_{ij} + \int_{-\infty}^\infty d\omega h'_{ij}(\omega) + \int_{-\infty}^\infty d\omega_2 \int_{-\infty}^\infty d\omega_1 h''_{ij}(\omega_1, \omega_2).
\end{align}
That is, the final expression can be written as a sum over a constant term ($h_{ij}$) and a single and double frequency integral over functions $h'_{ij}(\omega)$ and $h''_{ij}(\omega_1, \omega_2)$, respectively. Here, we use a convention that all the frequency integrals pass any pole on the real line from above (see \cref{fig_contour_avoiding_poles}, for example). Since our results are for general odd $J(\omega)$, these expressions cannot be reduced any further. Our \textit{Mathematica} files also provide numerical code to evaluate these integrals for arbitrary odd $J(\omega)$ as well as code to benchmark our results against HEOM and calculations done specifically for Drude cutoff (see \cref{sec_numerical_verification}).
The supplementary material also contains calculations for all the second order corrections to the SS, its equivalence to the corresponding mean force Gibbs state (MFGS) \cite{cresser2021weak} values (reported in our earlier work \cite{PhysRevB.111.115423}), as well as the analytical expression for the 4th order corrections to the SS coherences.

\subsection{Application to the Double-Quantum-Dot System}
\label{sec_dqd}

\begin{figure*}[htbp!]
    \centering
    
    \begin{subfigure}[b]{0.32\linewidth}
        \centering
        \includegraphics[width=\linewidth]{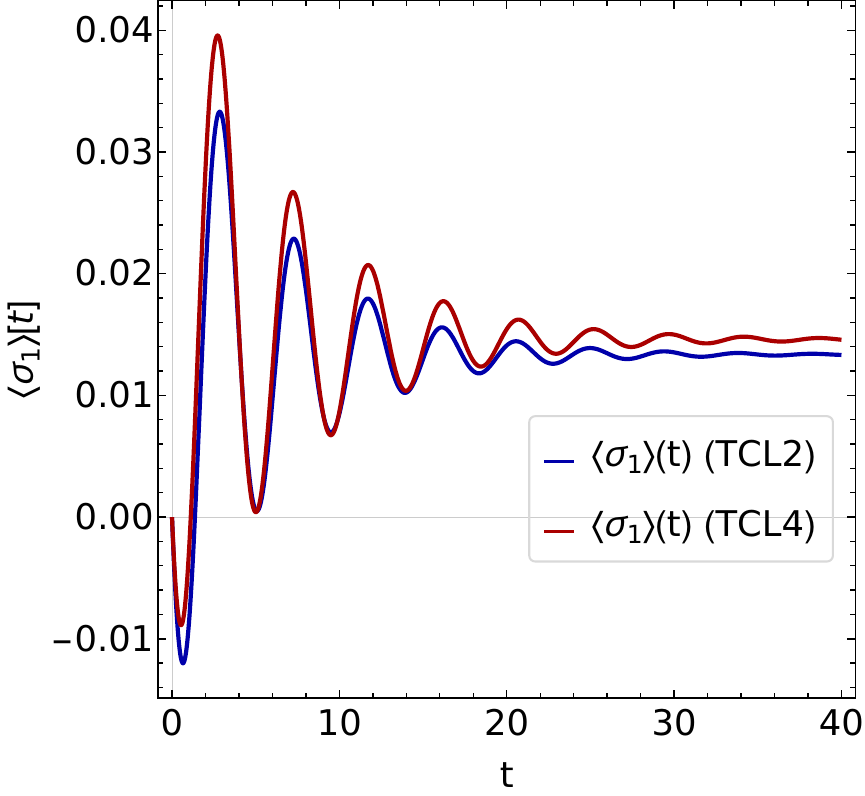}
        \caption{$\langle\hat{\sigma}_1(t)\rangle$}
        \label{fig_sigma1}
    \end{subfigure}
    \hfill 
    \begin{subfigure}[b]{0.32\linewidth}
        \centering
        \includegraphics[width=\linewidth]{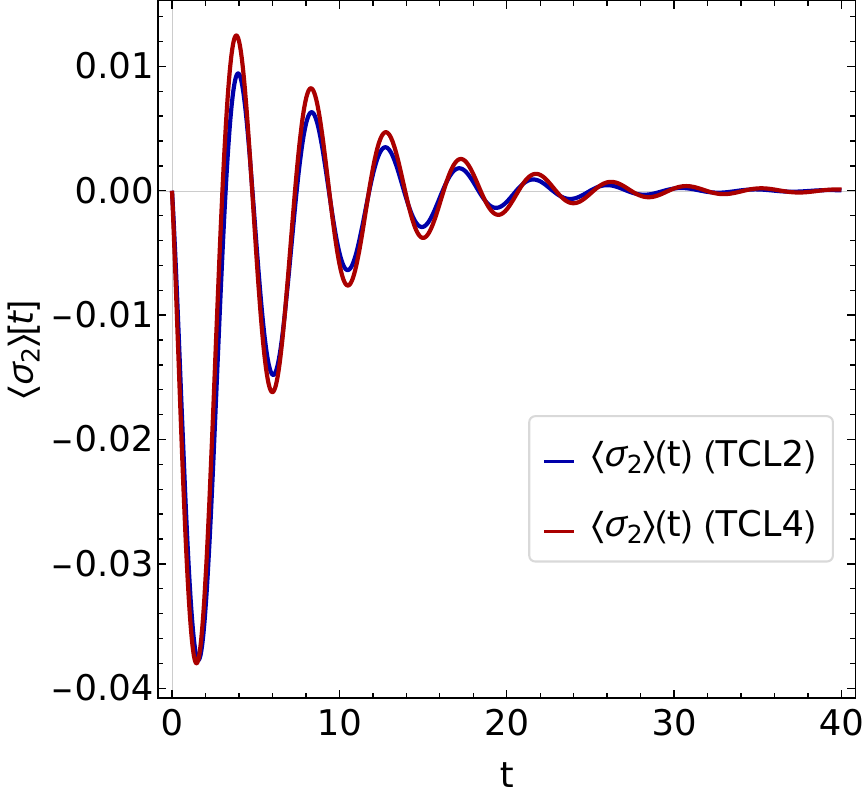}
        \caption{$\langle\hat{\sigma}_2(t)\rangle$}
        \label{fig_sigma2}
    \end{subfigure}
    \hfill 
    \begin{subfigure}[b]{0.32\linewidth}
        \centering
        \includegraphics[width=\linewidth]{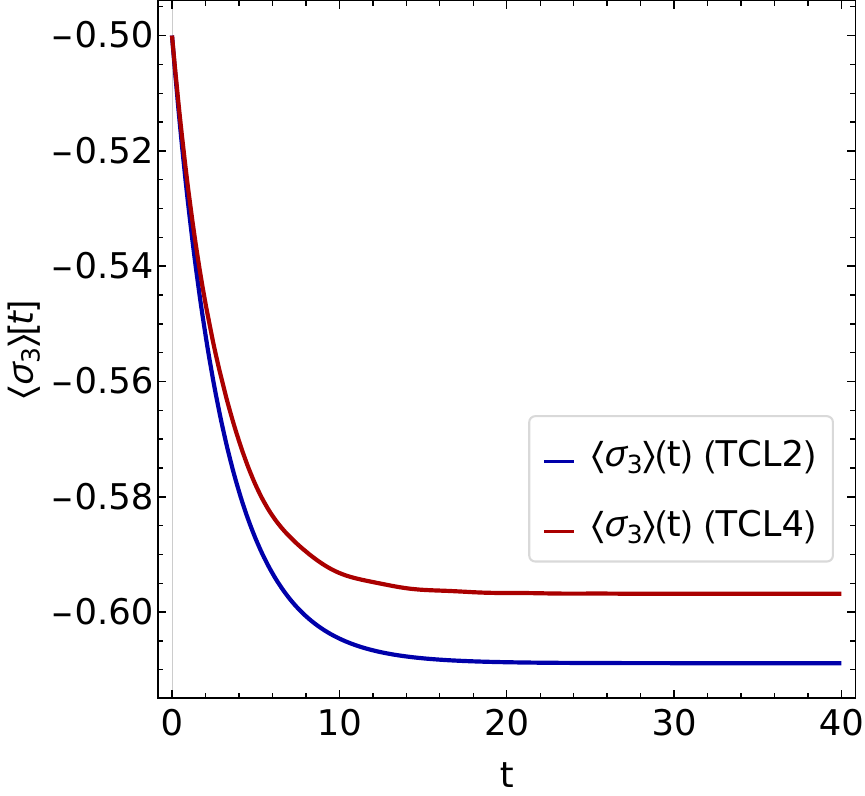}
        \caption{$\langle\hat{\sigma}_3(t)\rangle$}
        \label{fig_sigma3}
    \end{subfigure}
    \caption{Evolution of Pauli matrix expectation values (a) $\langle\hat{\sigma}_1(t)\rangle$, (b) $\langle\hat{\sigma}_2(t)\rangle$, and (c) $\langle\hat{\sigma}_3(t)\rangle$ for a DQD system using TCL2 (blue lines) and TCL4 (red lines) master equations. The model parameters are: $\epsilon = 1$, $t_c = 0.5$, $\gamma  \lambda ^2 = 0.4$, $\beta = 1$, $\omega _{\text{max}} = 1$ and $\omega_c = 1$. The initial values of the expectation values of these operators are chosen to be: $\left\langle \hat{\sigma}_1\right\rangle = 0$, $\left\langle \hat{\sigma}_2\right\rangle = 0$, $\left\langle \hat{\sigma} _3\right\rangle = -0.5$.}
    \label{fig:dqd_dynamics}
\end{figure*}

With the full asymptotic TCL4 generator $F^{(4)}$ at hand, we can now investigate the temporal dynamics of the SBM beyond the TCL2 approximation by solving
\begin{equation}
    \frac{d}{dt}\vec{v}(t) = (F^{(0)} + \lambda^2 F^{(2)} + \lambda^4 F^{(4)}) \vec{v}(t). \label{eq:TCL4_dynamics_eq}
\end{equation}
Here, we use this TCL4-ME to study the dynamics of a semiconductor DQD system, which can be effectively modeled by the SBM \cite{purkayastha2020tunable, PhysRevB.111.115423}. The SBM parameters are related to the DQD physical parameters (detuning $\epsilon$ and inter-dot tunneling $t_c$) as \cite{purkayastha2020tunable}:
\begin{align}
    a_{1} &= \frac{2 t_{c}}{\Omega},\\
    a_{3} &= \frac{\epsilon}{\Omega},\\
    \Omega^2 &= \epsilon^2 + 4 t_c^2.
\end{align}
The DQD is coupled to a phononic bath, described by the spectral density \cite{purkayastha2020tunable}:
\begin{equation}
  J(\omega) = \gamma \omega \, \left[1-\text{sinc}\left(\frac{\omega}{\omega_c}\right) \right]\, \exp\left\{-\frac{\omega^2}{2\: \omega_{\text{max}}^2}\right\}, \label{eqn_J_DQD}
\end{equation}
where, we have $\text{sinc}(x) \equiv \sin(x)/x$. This spectral density effectively describes the bulk acoustic phonons in Gallium Arsenide (GaAs) DQDs \cite{PhysRevLett.104.036801, colless2014raman, PhysRevB.97.035305, purkayastha2020tunable}. The parameter $\omega_{\text{max}}$ serves as the upper cut-off frequency, while $\omega_{c} = c_{s}/d$, where $c_{s}$ is the speed of sound in the substrate and $d$ is the inter-dot distance. The term $\lambda^2 \gamma$ dictates the coupling strength between the quantum dots and their phonon environment.
Also, we note that since this spectral density (\cref{eqn_J_DQD}) happens to be an odd function, our TCL4 results are directly applicable here.

\cref{fig:dqd_dynamics} plots the dynamics of a DQD system over time using TCL2 and TCL4. \cref{fig_sigma1}, \cref{fig_sigma2} and \cref{fig_sigma3} show corrections to $\braket{\hat{\sigma}_1(t)}$, $\braket{\hat{\sigma}_2(t)}$ and $\braket{\hat{\sigma}_3(t)}$ dynamics, respectively, for the initial state given as: $\left\langle \hat{\sigma} _1(0)\right\rangle = 0,\; \left\langle \hat{\sigma} _2(0)\right\rangle = 0,$ and $\left\langle \hat{\sigma} _3(0)\right\rangle = -0.5$.
The remaining model parameters are chosen to be: $\epsilon = 1,\; t_c = -0.5,\; \gamma  \lambda ^2 = 0.4,\; \beta = 1,\; \omega _{\text{max}} = 1$ and $\omega_c =1$. 

We note that the 4th order correction to the SS value of $\braket{\hat{\sigma}_2(t)}$ is zero, while the same is not true for the corresponding value of $\braket{\hat{\sigma}_1(t)}$. This is because for the present choice of the SBM model, the SS value of $\braket{\hat{\sigma}_2(t)}$ is zero at all orders in the perturbation theory, as noted in our previous work \cite{PhysRevB.111.115423}.
We conclude that TCL4 (compared to TCL2) provides corrections to the dynamics and SS of the DQD system that can become relevant in certain experimental conditions.

\subsection{Non-Markovian Effects with TCL4}
\label{sec_non_markovian_effects}

\begin{figure*}[htbp!]
  \centering
  \begin{subfigure}[b]{0.31\linewidth}
    \centering
    \includegraphics[width=\linewidth]{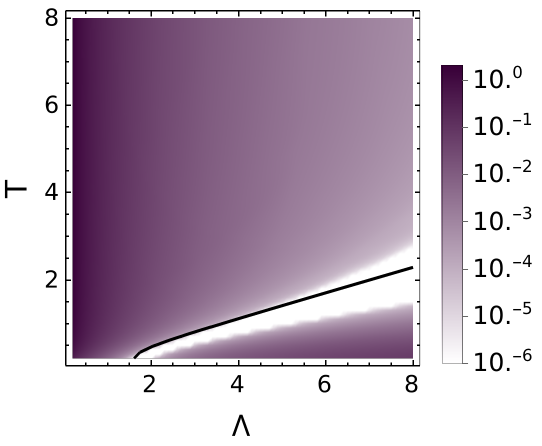}
    \caption{BLP measure using TCL2}
    \label{fig_blp_tcl2}
  \end{subfigure}
  \hfill 
  \begin{subfigure}[b]{0.31\linewidth}
    \centering
    \includegraphics[width=\linewidth]{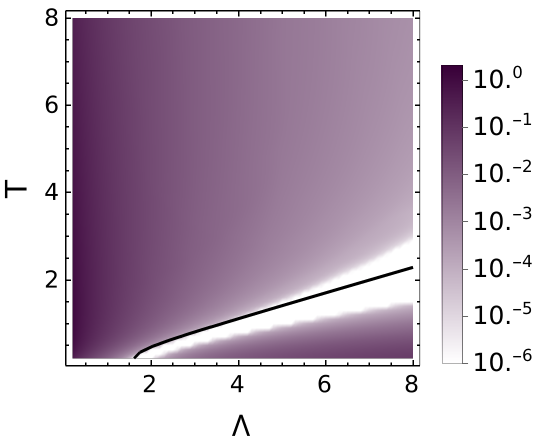}
    \caption{BLP measure using TCL4}
    \label{fig_blp_tcl4}
  \end{subfigure}
  \hfill 
  \begin{subfigure}[b]{0.34\linewidth}
    \centering
    \includegraphics[width=\linewidth]{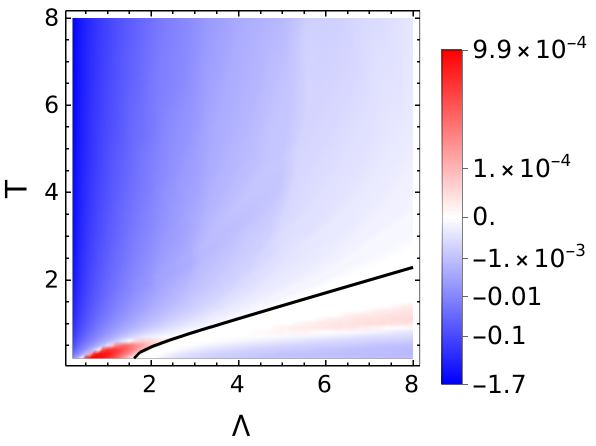}
    \caption{Difference: $N(\Phi_{\text{TCL4}})-N(\Phi_{\text{TCL2}})$}
    \label{fig_blp_diff}
  \end{subfigure}
  \caption{These figures plot the BLP measure ($N(\Phi)$) as logarithmic color plot calculated using (a) TCL2 and (b) TCL4 as a function of $\Lambda $ and $T$ for values of these parameters varying from $0.2$ to $8$. Part (c) plots the difference between these 2 quantities ($N(\Phi_{\text{TCL4}})-N(\Phi_{\text{TCL2}})$). The black line denotes the Markovian regime dictated by the resonance condition (see \cref{eqn_markovian_resonance_regime}). The maximization of the BLP measure was performed over $400$ antipodal state pairs uniformly chosen over the surface of the Bloch sphere. The remaining model parameters are fixed as $\Omega = 1.6$, $\gamma = 0.01$, and $\theta = \frac{\pi}{2}$.}
  \label{fig_NM_calcs_anti_p}
\end{figure*}

\begin{figure}[htbp!]
  \centering
  \includegraphics[width=1\linewidth]{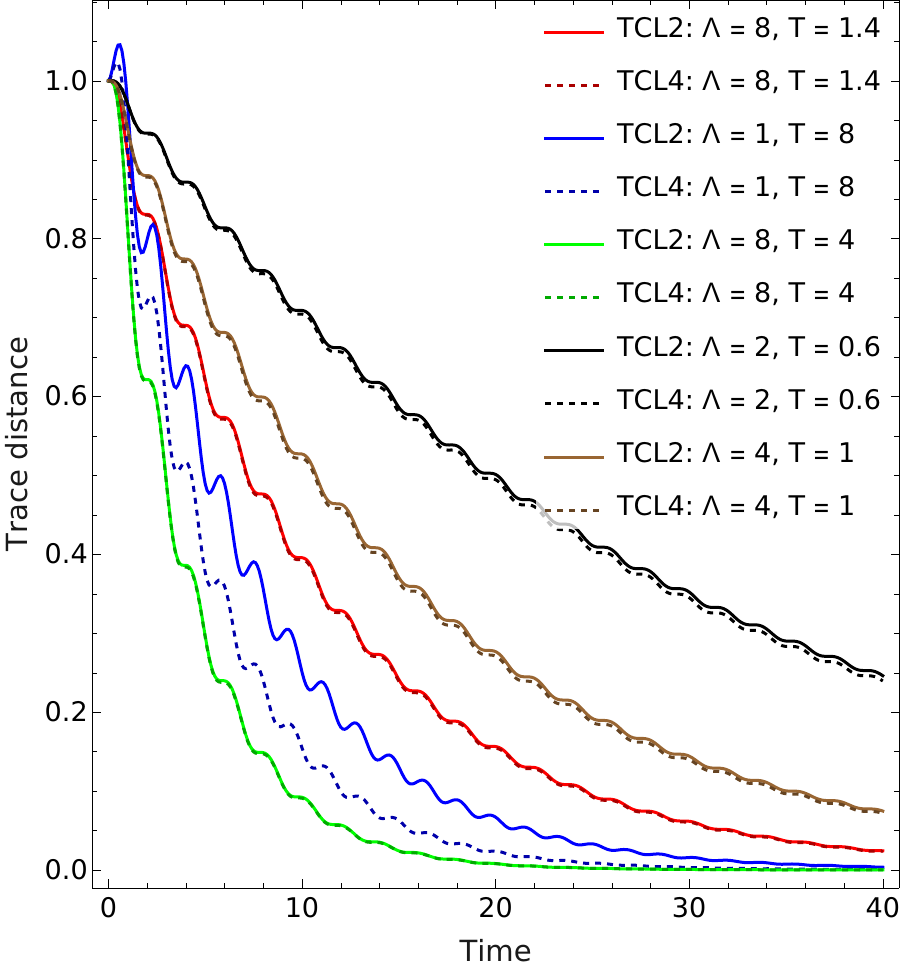}
  \caption{The trace distance between two states (given as $\vec{v}_1(0) = (1,1,0,0)$ and $\vec{v}_2(0) = (1,-1,0,0)$ in Bloch vectorized form (\cref{eqn_vectorization_of_qubit_state})) is plotted as a function of time for the TCL2 and TCL4  master equation. All the model parameters in this figure are the same as those in \cref{fig_NM_calcs_anti_p}.}
  \label{fig_tracedistance}
\end{figure}

In this section, we quantify the non-Markovianity of the TCL2 and TCL4 dynamics of the SBM using the Breuer-Laine-Piilo (BLP) measure \cite{breuer2009measure, clos2012quantification} for Ohmic spectral density with Drude cutoff, given as
\begin{align}\label{eqn_J_d_omega}
    J_D(\omega) = \frac{\gamma  \Lambda ^2 \omega }{\Lambda ^2+\omega ^2}.
\end{align}
Here, $\gamma$ is the coupling constant and $\Lambda$ is the environment cutoff frequency.

The BLP measure is derived based on the trace distance between two quantum states, which describes the distinguishability of these two states. The trace distance between two states is defined as \cite{nielsen2010quantum}
\begin{align} \label{eqn_trace_distance_defn}
    D(\hat{\rho}_1, \hat{\rho}_2 )= \frac{1}{2} \Tr|\hat{\rho}_1 - \hat{\rho}_2|,
\end{align}
where the modulus of an  operator $\hat{X}$ is $\left|\hat{X}\right| = \sqrt{\hat{X}^\dagger \hat{X}}$. For Markovian dynamics, the trace distance monotonically decreases over time, which indicates a continuous loss of distinguishability due to information backflow from the system to the environment. Any increase in trace distance at some time $t>0$  is considered a signature of non-Markovianity, indicating backflow of information from the environment to the system.

Assuming that the dynamics of the open quantum system is described by a completely positive trace preserving (CPTP) map $\Phi_{t}$, the degree of non-Markovianity of dynamics is defined by \cite{breuer2009measure}
\begin{equation}
 \mathcal{N}(\Phi) = \max_{\hat{\rho}_{1,2}(0)} \int_{\sigma > 0} dt\, \sigma(t, \hat{\rho}_{1,2}(0)),
 \label{eqn_blp_measure}
\end{equation}
where $\sigma(t, \hat{\rho}_{1,2}(0)) = \frac{d}{dt} D[\Phi_t(\hat{\rho}_{1}(0)), \Phi_t(\hat{\rho}_{2}(0))]
$. A non-zero value of $\mathcal{N}(\Phi)$ indicates backflow of information from the environment to the system. Since the measure $\mathcal{N}(\Phi)$ involves a maximization over all pairs of initial states, it quantifies the maximum possible information backflow. 

Here, we calculate $\mathcal{N}(\Phi)$ for different choices of temperature $T$ and cutoff parameter $\Lambda$ for the SBM \cite{clos2012quantification}.
We note that the BLP measure was originally derived for CPTP maps. Since the time-dependent TCL$2n$ is non-CP up to $O(\lambda^{2n})$ and is also not guaranteed to be CP-divisible, the asymptotic TCL$2n$ ME is also going to be non-CP. Yet, we extend the original motivation of the BLP measure, related to the revival of distinguishability of two states as a signature of information backflow, to use it to quantify non-Markovianity even in the case of the asymptotic TCL generators \cite{clos2012quantification}. We are also justified in doing this because the asymptotic TCL$2n$ ME is the long-time limit of the full TCL$2n$ ME, which is a physically derived approximate CPTP map.

\cref{fig_blp_tcl2} and \cref{fig_blp_tcl4} plot the value of $\mathcal{N}(\Phi)$ as logarithmic color plot for TCL2 and TCL4, respectively, for the values of $\Lambda$ and $T$ ranging from $0.2$ to $8$ each. The remaining model parameters in this figure are fixed as: $\Omega = 1.6$, $\gamma = 0.01$, and $\theta = \frac{\pi}{2}$.

As noted by Clos et al. \cite{clos2012quantification}, smaller values of $T$ and $\Lambda$ correspond to non-Markovian dynamics, except around the region given by the curve
\begin{align}\label{eqn_markovian_resonance_regime}
    \Lambda(T) &= \Omega  \sqrt{\frac{T \sinh \left(\frac{\Omega }{T}\right)+\Omega }{T \sinh \left(\frac{\Omega }{T}\right)-\Omega }},
\end{align}
where resonance condition between the system and environment gives rise to a Markovian regime. This Markovian regime is indicated as the white region in \cref{fig_blp_tcl2} and \cref{fig_blp_tcl4}.

For each data point in \cref{fig_blp_tcl2} and \cref{fig_blp_tcl4}, the maximization in the BLP measure (\cref{eqn_blp_measure}) was carried over $400$ sets of antipodal state pairs chosen uniformly from the surface of the Bloch sphere. See \cref{Bloch_parametrization} for further technical details and discussions.

\cref{fig_blp_diff} plots the difference between these two plots ($\mathcal{N}(\Phi_{TCL4}) - \mathcal{N}(\Phi_{TCL2})$) for the same range of $\Lambda$ and $T$. The blue region in \cref{fig_blp_diff}, which spans the majority of the parameter space considered in the figure, indicates that TCL2-ME \textit{overestimates} the non-Markovianity, as compared with TCL4-ME (and hence the exact result, since we are in a perturbatively weak coupling regime). On the other hand, we can also see regions, marked as red, where TCL2 \textit{underestimates} the non-Markovianity, as compared with the exact result, by a much smaller magnitude.

\cref{fig_tracedistance} plots the trace distance dynamics for initial vectorized states (\cref{eqn_vectorization_of_qubit_state}) given as $\vec{v}_1(0) = (1,1,0,0)$ and $\vec{v}_2(0) = (1,-1,0,0)$ for different choices of $\Lambda$ and $T$, where the other model parameters are kept the same as in \cref{fig_NM_calcs_anti_p}. Note that the choice of $\Lambda$ and $T$ is such that it samples curves from the red, blue, and white regions in \cref{fig_blp_diff}. Studying the trace distance dynamics in these regions can help us understand their distribution within the parameter space.

The dark blue region in \cref{fig_blp_diff} tend to exist in high non-Markovian regimes, where the correction from TCL4 is significant enough to cause a faster equilibration of the system (see $(\Lambda = 1, T = 8)$ curves in \cref{fig_tracedistance}), giving rise to faster decay and hence decreased revival of the distinguishability. Hence, the region near the white Markovian regime in \cref{fig_blp_diff} becomes a necessary but not sufficient condition for the existence of the red region in this figure. Here, the TCL4 correction is not strong enough to significantly speed up the equilibration, yet can, in some parameter regimes, give rise to a positive but small contribution to the revival of distinguishability (for example, see $(\Lambda = 8$, $T=1.4)$  and $(\Lambda = 2$, $T=0.6)$ curves in \cref{fig_tracedistance}). We can contrast this with the $(\Lambda = 8$, $T=4)$ curve in \cref{fig_tracedistance}, where TCL4 does not significantly speed up the equilibration, yet TCL2 \textit{overestimates} the non-Markovianity.

For completeness, in \cref{fig_tracedistance}, we also provide the curves corresponding to the white Markovian region in \cref{fig_blp_diff} (the $\Lambda = 4$, $T=1$ curves). For this curve, the trace distance dynamics goes through a monotonic decay for both TCL2 and TCL4.

\section{Numerical Verification and Benchmarking}
\label{sec_numerical_verification}

In this section, we perform 2 types of numerical verifications and benchmarking of our results:
\begin{enumerate}
    \item Consistency checks against TCL4 calculations done for a specific spectral density.
    \item Benchmarking against the numerically exact HEOM method.
\end{enumerate}

\subsection{Consistency for Ohmic spectral density with Drude Cutoff}
\label{sec_drude_verification}
\begin{figure}[htbp!]
  \centering
  \includegraphics[width=\linewidth]{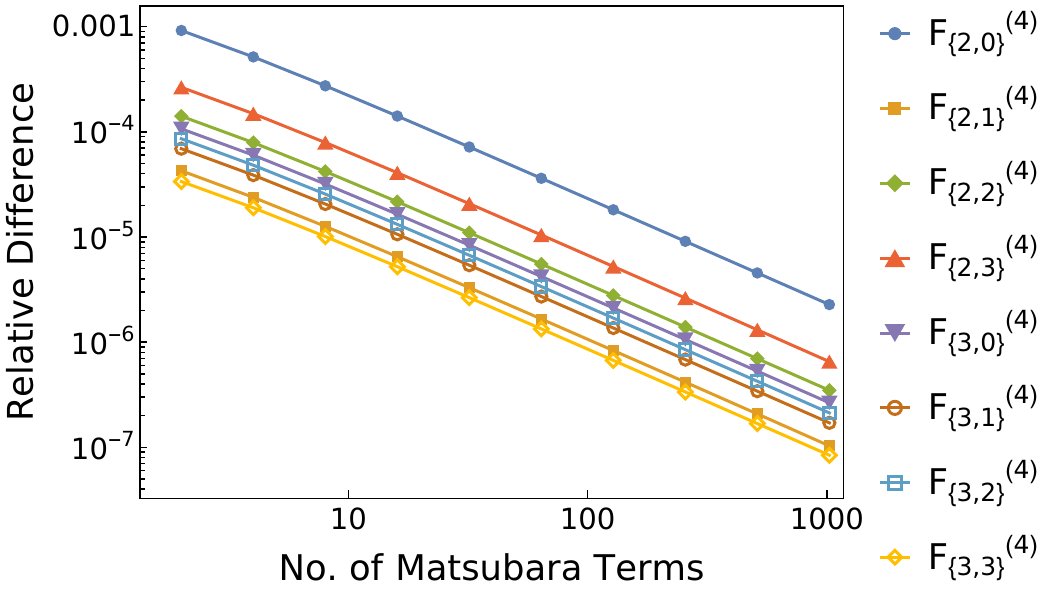}
  \caption{Relative difference between matrix elements of the asymptotic TCL4 generator ($F^{(4)}$) for Ohmic spectral density with Drude cutoff, calculated using our general result (valid for all odd spectral densities) and specialized calculations done specifically for the current choice of the spectral density, as a function of the number of Matsubara terms considered in the latter calculation. All free parameters of the model were randomly chosen (in weak coupling limit) and are specifically given as: $\gamma = 3.58\times 10^{-3}$, $\Lambda = 2.07\times 10^{-1}$, $\Omega = 1.91\times 10^{-1}$, $\beta = 3.15\times 10^{-1}$, $\theta = 1.83\times 10^{-1}$. Note that both the $x$ and $y$ axes on this plot are in log scale.}
  \label{fig_tcl4_verification}
\end{figure}
For an Ohmic spectral density with Drude cutoff (see \cref{eqn_J_d_omega}), closed-form expressions for the two-point correlation functions $\nu(t)$ and $\eta(t)$ (\cref{eqn_eta_expr,eqn_nut_expr}) are available as \cite{breuer2002theory}:
\begin{align}\label{eqn_drude_eta}
    \eta_D(t) &= -\frac{1}{2} \pi  \gamma  \Lambda ^2 e^{-t \Lambda},\\
    \nu_D(t) &= \sum_{n = -\infty}^\infty \frac{\pi  \gamma  \Lambda ^2 \left(\Lambda  e^{-\Lambda t}-| \mu_n|  e^{- t | \mu_n| }\right)}{\beta  \left(\Lambda ^2-|\mu_n| ^2\right)}\label{eqn_drude_nu}.
\end{align}
Here, $\mu_n = 2 \pi n/\beta$ denotes the $n$th Matsubara frequency.

Consequently, these closed-form expressions can be substituted directly into the time-domain integrals that define the TCL2 and TCL4 coefficients (see \cref{eqn_a334_triple_integral}) and evaluated in the long-time limit. This procedure circumvents the double $\omega$-integrals that were necessary in the more general treatment.
The final result is symbolically expressed as
\begin{align}
    X_{ij}(N) &= d_{ij} + \sum_{n = 1}^N d'_{ij} (n) + \sum_{n, m = 1}^N d''_{ij} (n, m)\\
    {F_D}_{ij}^{(4)} &= \lim_{N \to \infty} X_{ij}(N)
\end{align}
Here, ${F_D}_{ij}^{(4)}$ is the corresponding TCL4 generator for $J(\omega) = J_D(\omega)$ (\cref{eqn_J_d_omega}).
This means that the final expression can be written as a sum over a constant term ($d$) and a single and double infinite summation over functions $d'(n)$ and $d''(n, m)$, respectively.

Although we calculated the expression of $d$, $d'(n)$ and $d''(n, m)$ for all the TCL4 coefficients ${F_D}_{ij}^{(4)}$, we are not able to do these infinite summations in closed form for all the values of $\{i,j\}$ \cite{SupplementaryMaterial}.
Instead, we numerically do this summation for a large and finite $N$ to verify our general results. 

In \cref{fig_tcl4_verification}, we plot the relative difference, defined as
\begin{align}
    E_{\text{rel}} (N)&\equiv ({F'}_{ij}^{(4)} - X_{ij}(N))/{F'}_{ij}^{(4)},
\end{align}
as a function of $N$. Here ${F'}_{ij}^{(4)}$ represents the numerical value of the TCL4 generator, obtained from our general results, for $J(\omega) = J_D(\omega)$. The free parameters of the model were chosen randomly (in weak coupling limit) to be: $\gamma = 3.58\times 10^{-3},\; \Lambda = 2.07\times 10^{-1},\; \Omega = 1.91\times 10^{-1},\; \beta = 3.15\times 10^{-1},\; \theta = 1.83\times 10^{-1}$. Note that both the $x$ and $y$ axes on this plot are in log scale, and the relative difference for all coefficient matrices decays linearly. For some positive coefficient $m$, this implies a $1/N^m$ type of decay of the relative difference. Hence, this figure numerically indicates that the relative difference between these two quantities is at least $O(10^{-6})$ and decays off quickly when more Matsubara terms are considered.

We note that we can analytically evaluate the infinite summations for ${F_D}_{3,0}^{(4)}$ and ${F_D}_{2,0}^{(4)}$. The subsequent relative difference is found to be $O(10^{-12})$ (and can be further brought down by using higher precision numerical integration), providing further numerical support for the validity of our results \cite{SupplementaryMaterial}.

\subsection{Comparison with HEOM}
\label{sec_heom_verification}
\begin{figure}[htbp!]
  \centering
  \includegraphics[width=1 \linewidth]{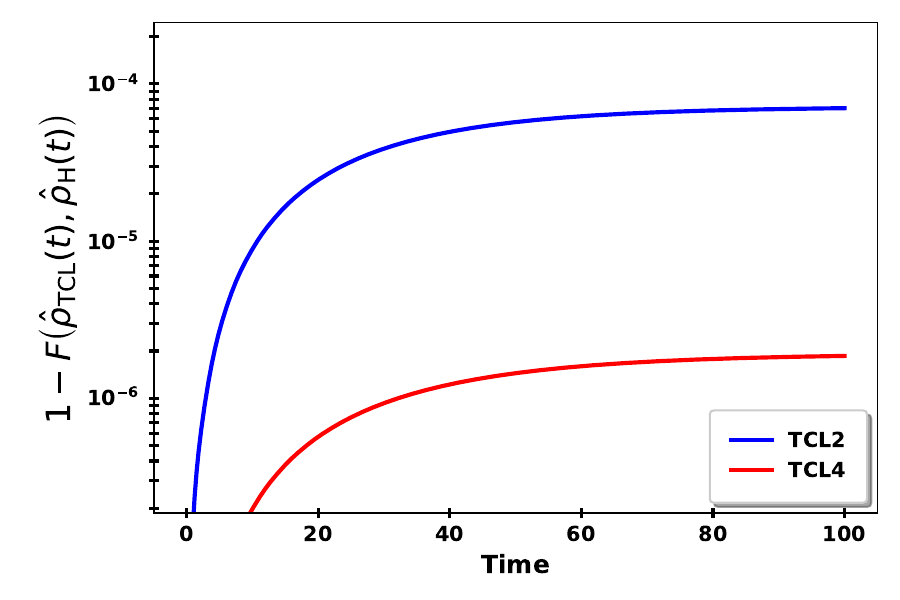}
  \caption{One minus the Fidelity of the system state evolved by HEOM and TCL-ME (blue line is TCL2 and red line is TCL4). Calculations are for an SBM with an Ohmic spectral density with Drude cutoff and with the model parameters given as $\gamma  \lambda ^2 = 0.02$, $\Lambda = 1$, $\Omega = 1.6$, $\beta = 1$, $\theta = 0.785$. For HEOM calculations, 32 Matsubara terms and 2 levels of hierarchy were retained.}
  \label{fig_fidelity_heom}
\end{figure}

The HEOM technique is a numerically exact approach for solving the dynamics of an open quantum system based on Feynman-Vernon
influence functional formalism \cite{doi:10.1143/JPSJ.58.101, PhysRevA.41.6676, doi:10.1143/JPSJ.75.082001,tanimura2020numerically}.
Here, we use the HEOM solver provided in the Qutip package \cite{JOHANSSON20121760, lambert2024qutip5quantumtoolbox} to benchmark the TCL2 and TCL4 dynamics for a generic SBM. 
The numerical implementation of HEOM comes with several precision parameters that need to be tuned to achieve the desired accuracy. For the present consideration, these include
\begin{enumerate}
    \item the number of Matsubara terms retained: These correspond to the number of the Matsubara terms to be retained in the expression of the two-point correlation function in \cref{eqn_drude_eta} and \cref{eqn_drude_nu}. A higher number of Matsubara terms need to be retained in the low temperature limit \cite{doi:10.1143/JPSJ.75.082001}.
    \item the number of hierarchical equations retained: The HEOM technique involves expressing the open system dynamics in terms of an infinite hierarchy of differential equations for the so called auxiliary density operators (ADOs), where the zeroth order term corresponds to the exact time dependent state of the system \cite{PhysRevA.41.6676}. If these coupled differential equations are truncated at some order, the system of differential equations becomes numerically tractable, since the initial state of all the ADOs is known.
\end{enumerate}

\Cref{fig_fidelity_heom} compares the dynamics predicted by TCL2 and TCL4-MEs with those obtained from HEOM for Ohmic spectral density with Drude cutoff (\cref{eqn_J_d_omega}).
The comparison is quantified by the fidelity, \cite{nielsen2010quantum}
\begin{align}
F(\hat{\rho}_{\text{TCL}}(t), \hat{\rho}_{\text{H}}(t)) = \Tr \sqrt{\sqrt{\hat{\rho}_{\text{H}}(t)} \hat{\rho}_{\text{TCL}}(t) \sqrt{\hat{\rho}_{\text{H}}(t)}},
\end{align}
between the system state evolved by the TCL ME ($\hat{\rho}_{\text{TCL}}$) and the HEOM state ($\hat{\rho}_{\text{H}}$).
The model parameters in the figure are: $\gamma  \lambda ^2 = 2.\times 10^{-2},\; \Lambda = 1,\; \Omega = 1.6,\; \beta = 1,\; \theta = 7.85\times 10^{-1}$, while at the HEOM side, we retained 32 Matsubara terms (\cref{eqn_drude_eta} and \cref{eqn_drude_nu}) and 2 levels of hierarchies. We checked the robustness of the results in this plot by going to higher Matsubara terms and hierarchy depths (128 and 4, respectively). We note that closeness between the TCL4 and HEOM dynamics increases as we increase the precision of the HEOM calculations, while the same is not true for TCL2 dynamics.

Note that since our TCL4 result is applicable in the asymptotic limit ($ t \gg \tau_B$), the above calculations were performed in this same limit. That is, we initialized a state $\hat{\sigma}(0) = \ket{+}\bra{+}$, then calculated a time $\tau \gg \{\Lambda^{-1}, \beta\}$,  (here, $\tau = 60$), and then defined $\hat{\rho}_H(t) \equiv \hat{\sigma}(t + \tau)$ and initialized the TCL state as $\hat{\rho}_{TCL}(0) = \hat{\sigma}(\tau)$.

For ease of visualization, \Cref{fig_fidelity_heom} plots $1 - F(\hat{\rho}_{\text{TCL}}(t), \hat{\rho}_{\text{H}}(t))$ for both TCL2 and TCL4. We conclude that TCL4-ME maintains a significantly higher fidelity with the HEOM results compared to TCL2-ME.

\section{Discussion and Conclusion}
\label{sec_conclusion}

In this work, we have derived the complete analytical form of the asymptotic fourth-order time-convolutionless (TCL4) generator, $F^{(4)}$, for the generic spin-boson model, under the assumption that the environmental spectral density, $J(\omega)$, is an odd function of frequency. The key results of this work are:
\begin{itemize}
    \item We have simplified the final expressions for $F^{(4)}$ in terms of a single or double integral over frequencies (see \cref{eqn_symbolic_tcl4_solution} and \cref{appendix_supplementary_material}), which cannot be further simplified without knowing the explicit form of the spectral density.
    \item This has enabled us to evaluate the expressions for the 4th-order correction to the SS coherences for the generic SBM (see \cref{appendix_supplementary_material}).
    \item The full $F^{(4)}$ generator enables the calculation of the complete temporal dynamics of the SBM up to $O(\lambda^4)$ in the SE coupling strength $\lambda$. For example, application to a DQD system demonstrates that TCL4 dynamics provides corrections to the dynamics that may be of experimental relevance under some parameter regimes (see \cref{sec_dqd}).
    \item We report that for Ohmic spectral density with Drude cutoff, over a large range of temperature $T$ and cutoff frequency $\Lambda$, TCL2 overestimates the non-Markovianity of the dynamics (see \cref{fig_tracedistance}).
    \item The analytical expressions for the $F^{(4)}$ generator were rigorously validated through consistency checks against specialised calculations for the Ohmic spectral density with Drude cutoff and by benchmarking the predicted dynamics against the numerically exact HEOM method.
\end{itemize}
All the results mentioned in this work, including symbolic derivations, final expressions, numerical evaluation for specific parameters, and plots, etc., are publicly available in their complete form as the accompanying Supplementary material \cite{SupplementaryMaterial}.
These results can serve as a valuable tool for theoretical investigations of a wide array of physical systems and phenomena modelled by the generic SBM \cite{gilmore2006criteria, Gilmore_2005, gilmore2008quantum, porras2008mesoscopic, cheche2001dynamics, merkli2013electron, magazzu2018probing}. It provides a means to explore dynamics in regimes where TCL2 is insufficient, without resorting to computationally intensive methods like HEOM and tensor network approaches (see, for example, \cite{chen2025benchmarking} regarding this `computational intensiveness').

Future research directions could include efforts to relax the assumption of an odd spectral density that we have assumed in our work, which would broaden the applicability of the present result, albeit at the cost of possible increased analytical complexity. Exploring the extension to even higher orders (e.g., TCL6 and beyond (see \cite{PhysRevA.111.042214})) could be considered, though the analytical challenges are formidable. 
Furthermore, at finite temperature, the correspondence between the $O(\lambda^4)$ corrections to SS coherences (derived in this work) and the corresponding MFGS quantity can also be checked, potentially shedding light on the conditions for their equivalence or divergence (the latter being indicated in \cite{crowder2024invalidation} at zero temperature), a topic of ongoing interest.

\bibliographystyle{IEEEtran}
\bibliography{./citation}

\clearpage  

\appendix
\crefname{section}{Appendix}{Appendices}
\Crefname{section}{Appendix}{Appendices}
\makeatletter

\section{Parametrization on the Bloch sphere}\label{Bloch_parametrization}
\begin{figure*}[htbp!]
  \centering
  \begin{subfigure}[b]{0.31\linewidth}
    \centering
    \includegraphics[width=\linewidth]{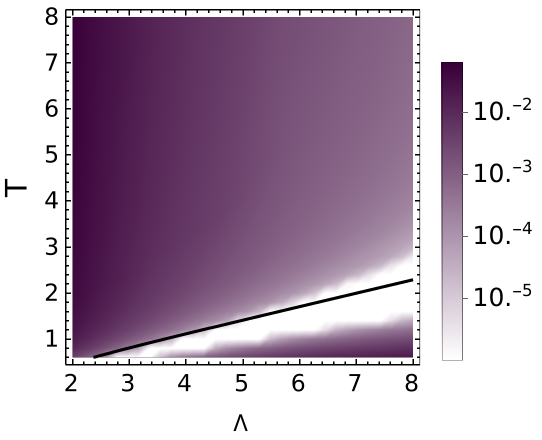}
    \caption{BLP measure using TCL2}
    \label{fig:gen_tcl2}
  \end{subfigure}
  \hfill 
  \begin{subfigure}[b]{0.31\linewidth}
    \centering
    \includegraphics[width=\linewidth]{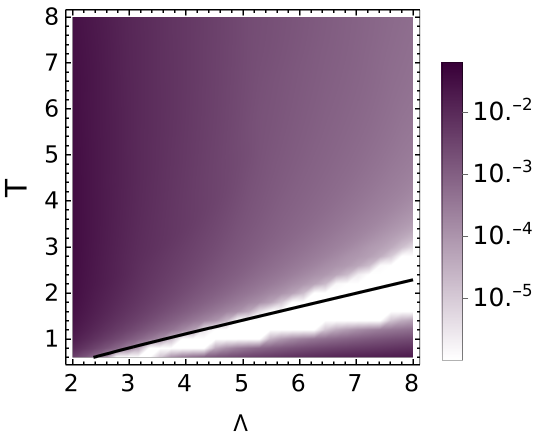}
    \caption{BLP measure using TCL4}
    \label{fig:gen_tcl4}
  \end{subfigure}
  \hfill 
  \begin{subfigure}[b]{0.34\linewidth}
    \centering
    \includegraphics[width=\linewidth]{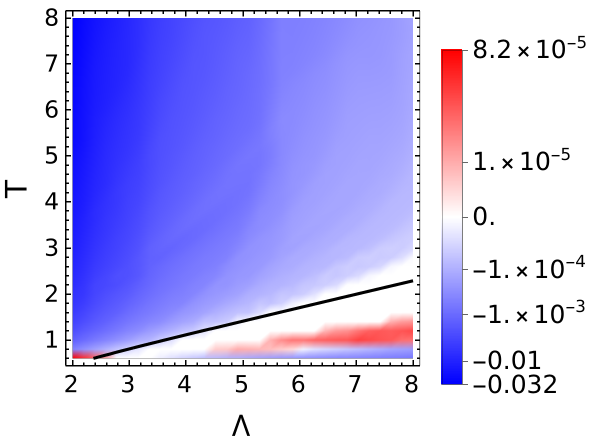}
    \caption{Difference: $N(\Phi_{\text{TCL4}})-N(\Phi_{\text{TCL2}})$}
    \label{fig:gen_diff}
  \end{subfigure}
  \caption{This figure is analogous to \cref{fig_NM_calcs_anti_p}, the only difference being that over here, the maximization of the BLP measure (\cref{eqn_blp_measure}) was performed over $4096$ state pairs (not necessarily antipodal, as was the case in \cref{fig_NM_calcs_anti_p}) uniformly chosen over the surface of the Bloch sphere. Also, over here, the $\Lambda$ and $T$ parameters range from $2$ to $8$ and $0.6$ to $8$, respectively, since for lower $\Lambda$ and $T$ values, maximization over a large number of state ensembles becomes numerically expensive as the relaxation time, $t_{max}$, tends to increase significantly.}
  \label{fig_NM_calcs}
\end{figure*}

In \cref{fig_NM_calcs_anti_p}, to generate Bloch vectors that are uniformly distributed over the surface of the Bloch sphere, we use the following parametrisation.  Let $ u, v \in [0,1]$ be independent random variables drawn from a uniform distribution. The Cartesian component  of the corresponding Bloch vector is then given by: 
\begin{align*}
x(u,v) &= \cos(2\pi u) \cdot \sin\left( \cos^{-1}(2v - 1) \right), \\
y(u,v) &= \sin(2\pi u) \cdot \sin\left( \cos^{-1}(2v - 1) \right), \\
z(v)   &= \cos\left( \cos^{-1}(2v - 1) \right).
\end{align*}
This mapping ensures that the generated points are uniformly distributed over the surface of the sphere, thereby producing a uniform distribution of Bloch vectors on the Bloch sphere.

To calculate the non-Markovianity measure $\mathcal{N}(\Phi)$ in \cref{fig_NM_calcs_anti_p}, we uniformly generate $20$ values of variables $u$ and $v$, each, from their respective domains, giving rise to a set of $400$ states, $[x(u,v), y(u,v), z(v)]$, on the Bloch surface. For each such state, the corresponding antipodal state is chosen as its pair, and the corresponding BLP calculation is done as explained in Sec. \ref{sec_non_markovian_effects}. The trace distance between these antipodal state pairs is evolved until it decays close enough to zero, as can be seen in \cref{fig_tracedistance} for some of the curves. For numerical optimization, this time interval, $t_{max}$, for which the BLP integral (\cref{eqn_blp_measure}) needs to be calculated, is determined for each value of $\Lambda$ and $T$.
For given values of $\Lambda$ and $T$, let $\hat{\rho}_{\Lambda,T}(0)$ be the state farthest away from the corresponding Gibbs state in terms of trace distance. Then $t_{max}$ is determined as the smallest time such that $D(\hat{\rho}_{\Lambda,T}(t_{max}), \hat{\rho}_{eq}) < \epsilon$, where $\hat{\rho}_{\Lambda,T}(t_{max})$ is the initial state evolved under TCL2, $\hat{\rho}_{eq}$ is the SS of TCL2 and $\epsilon$ is a small threshold, chosen to be equal to $0.001$. We note that the qualitative features in \cref{fig_NM_calcs_anti_p} were found to be robust under variation of $\epsilon$ and other numerical precision parameters.

Note that for CP maps, antipodal points are known to maximize the BLP measure \cite{wissmann2012optimal}.
Even though in the present work, we are not dealing with CP maps, we restrict ourselves to antipodal points for numerical simplification.
But such a restriction may come at a cost to the accuracy of the corresponding results. To check for this, \cref{fig_NM_calcs} reports the plots done without restricting ourselves to antipodal points, albeit at a lower precision and range of $\Lambda$ and $T$. Here, the state pairs $[x(u,v), y(u,v), z(v)]$ and $[x(u_1,v_1), y(u_1,v_1), z(v_1)]$ are again chosen uniformly from the Bloch surface, where the variables $u, v, u_1, v_1 \in [0,1]$ now vary over $8$ uniformly chosen values in their domain. This results in a total of $8^4 = 4096$ pairs of quantum states sampled from the surface of the Bloch sphere, over which the BLP measure was maximized. All other parameters in \cref{fig_NM_calcs} are the same as those in \cref{fig_NM_calcs_anti_p}. We note that \cref{fig_NM_calcs} qualitatively reproduces all features of \cref{fig_NM_calcs_anti_p} essential for our results.

\section{How to use the Supplementary material}
\label{appendix_supplementary_material}
All the main results of this work are available in `\verb|TCL4DynamicsSpinBosonResults.nb|' file \cite{SupplementaryMaterial}.

\subsection{TCL General Derivation}

`\verb|TCLIntegrandCalcs.wl|' starts from the expression of the TCL2 and TCL4 generators that can be found in \cite{breuer2001time}, and derives the matrix elements of the TCL4 and TCL2 generator in Bloch vectorized format. `\verb|TCL2GeneratorCalc.wl|' uses these results to evaluate the general TCL2 generator. Whereas, `\verb|TCL4TripleTimeIntegral.wl|' symbolically evaluates the triple time integrals of the TCL4 generator matrix elements. `\verb|TCL4OmegaIntegral.wl|' then evaluates the frequency integrals in the large time limit.

`\verb|DQDTCL4GeneratorEvaluation.wl|' reproduces the \cref{fig:dqd_dynamics} and can also be used to solve for dynamics for any other odd spectral density. For example, `\verb|TCL4GeneratorEvaluationForDrude.wl|' performs similar calculations for the Ohmic spectral density with Drude cutoff.

`\verb|O2SSCalcSpinBoson.nb|' analytically evaluates the 2nd and 4th order SS results.

\subsection{Verification of our results}

`\verb|TCL4DrudeGeneratorCalc.wl|' and `\verb|TCL2DrudeGeneratorCalc.wl|' do the TCL4 and TCL2 calculations for Ohmic spectral density with Drude cutoff. Subsequently, `\verb|TCL4Verification.wl|' verifies our general result with these Drude calculations and reproduces \cref{fig_tcl4_verification}. `\verb|TCL4HighPrecisionVerification.wl|' does the same verification at higher precision for $F_{20}^{(4)}$ and $F_{30}^{(4)}$.

`\verb|TCLVsHEOMFidelity.wl|' and `\verb|tcl_vs_heom.py|' together benchmark our results using HEOM. This is achieved like this: `\verb|TCLVsHEOMFidelity.wl|' does the calculations at the TCL side, dumps all the results into a text file in the same folder, which will be named `\verb|python_objects.txt|'. Then, `\verb|tcl_vs_heom.py|' has to be run, which will read these results from `\verb|python_objects.txt|' and do the HEOM side calculations using the qutip python package, hence comparing these two results and reproducing \cref{fig_fidelity_heom}.

`\verb|AntipodalTCL2NM.wl|' and `\verb|AntipodalTCL4NM.wl|' do the non-Markovianity calculations for antipodal states while `\verb|AntipodalSBMProjectNMPlot.wl|' plots the results (\cref{fig_NM_calcs_anti_p}). Similarly, `\verb|GeneralTCL2NM.wl|' and `\verb|GeneralTCL4NM.wl|' do the non-Markovianity calculations for general states while `\verb|PlotSBMProjectNM.wl|' plots the results (\cref{fig_NM_calcs}). Finally, `\verb|TraceDistanceDynamicsPlot.wl|' reproduces \cref{fig_tracedistance}.

\end{document}